\documentclass[aps,prb,twocolumn,superscriptaddress,amsmath,amssymb,longbibliography]{revtex4-2}

\usepackage{bm}
\usepackage{amsmath}
\usepackage{amssymb}
\usepackage{graphicx}
\graphicspath{{img_1/}}
\usepackage{xcolor}
\input{epsf}
\usepackage[english]{babel}
\usepackage{gensymb}
\usepackage{ulem}
\usepackage[colorlinks,citecolor=cyan,linkcolor=cyan,urlcolor=cyan,bookmarks=false,hypertexnames=true]{hyperref}

\usepackage{booktabs}  

\begin{document}

	\title{Rapid Determination of Nanodiamond Size Distribution and Impurity Concentration from Raman Spectra Using an Open Machine-Learning Toolbox
    }

% Open machine learning toolbox for fitting nanodiamond Raman spectra and extracting size distribution and impurity concentration \\     Rapid Determination of Nanodiamond Size Distribution and Impurity Concentration from Raman Spectra Using an Open Machine-Learning Toolbox 
    %Open machine-learning toolbox for fitting nanodiamond Raman spectra allowing determination of size distribution and impurities concentration
%1. Determination of the size distribution of nanodiamonds from Raman spectra by continuous bond-polarization model-based automatized optimization tool\\2. 
%    \\\re{3. (Machine Learning-inspired? -- for more hype and citations) Tool for nanodiamond ensembles Raman spectra interpretation (fitting, treatment)} 

	\author{Sergei V.~Koniakhin}
	\email{kon@ibs.re.kr}
	\affiliation{Center for Theoretical Physics of Complex Systems, Institute for Basic Science (IBS), Daejeon 34126, Republic of Korea}
	\affiliation{Basic Science Program, Korea University of Science and Technology (UST), Daejeon 34113, Republic of Korea}
    \author{Oleg I. Utesov}
	\affiliation{Center for Theoretical Physics of Complex Systems, Institute for Basic Science (IBS), Daejeon 34126, Republic of Korea}
    \author{Vitaly I. Korepanov }
	\affiliation{Institute of Microelectronics Technology RAS, 6 Academician Ossipyan str., Chernogolovka 142432, Russia}
    \author{Andrey~G.~Yashenkin}
    \affiliation{The Faculty of Physics of St. Petersburg State University, Ulyanovskaya 1, St. Petersburg 198504, Russia}
    \affiliation{Petersburg Nuclear Physics Institute NRC ``Kurchatov Institute'', Gatchina 188300, Russia}

	\begin{abstract}

        Ready-to-use numerical toolbox for nanodiamond Raman spectra calculation and fit is presented. The developed theoretical approach allows accounting for arbitrary nanoparticle size-distribution and the microscopic line broadening mechanisms for the optical phonons. The two tools for solving the inverse problem of the nanodiamond properties reconstruction using a known Raman spectrum are provided. The first one utilizes a dense neural network trained on a vast array of synthetic Raman spectra. The second approach is based on the stochastic Metropolis algorithm, which updates the ensemble parameters by small quantities, tending to the state with minimal error. Both methods are available thanks to the computationally instant elasticity theory-like model for optical phonon modes in diamond nanocrystals that accurately reproduces the results of the atomistic approaches. Using experimental Raman spectra for nanodiamonds prepared by various techniques, we tested our tools and observed a faithful agreement with the data as well as between the two methods. The open and documented software is accessible online  (\href{https://nanoraman.pythonanywhere.com/}{nanoraman.pythonanywhere.com}) and as a Python module (\href{https://github.com/KoniakhinSV/Nanoparticle\_Raman}{github.com/KoniakhinSV/Nanoparticle\_Raman}).

	\end{abstract}

	\maketitle


	\section{Introduction}
	
	Raman spectroscopy remains one of the most widely used, non-destructive probes of the structural and vibrational properties of carbon materials~\cite{ferrari2004raman,vul2013detonation}. It is particularly sensitive to nanoscale effects in diamond nanoparticles (nanodiamonds), where size quantization (also referred to as confinement), surface chemistry, and disorder jointly determine the observed lineshapes and peak positions~\cite{korepanov2017carbon,koniakhin2018raman,utesov2020lifetimes,koniakhin2020lifetimes,utesov2021effects}. The first-order diamond Raman line (located near 1333~cm$^{-1}$ in bulk diamond) typically experiences a size-dependent redshift and asymmetric broadening in nanoparticle ensembles. These phenomena have been broadly attributed to relaxation of the momentum selection rules and to contributions from phonons with finite wavevector brought about by finite crystallite size (the so-called phonon confinement picture)~\cite{richter1981onephonon,campbell1986effects}. Over the last decades, this phenomenological framework has been extended to numerous nanomaterials, linking Raman lineshapes to particle-size distributions (see, e.g., Refs.~\cite{arora2004phonon,gouadec2007raman,osswald2007phonon, korepanov2020localized}). An alternative approach to the nanoparticle size effect on the Raman spectra shape, explicitly coupling atomic contributions with various hybridization states, was proposed in Ref.~\cite{gao2019determination}.

	Despite its utility, the original naive phonon confinement model (PCM) suffers from conceptual and quantitative limitations when applied to real nanoparticle powders. These limitations include sensitivity to the assumed phenomenological confinement function~\cite{zi1997comparison}, neglect of anisotropic phonon dispersion and mode mixing~\cite{osswald2007phonon}, and an often uncontrolled treatment of surface and defect contributions that can produce spectral shifts and broadening comparable to size effects ~\cite{mochalin2009contribution}. The lack of physical motivation of the confinement function broadening controlled by the $\alpha$ parameter is the most significant issue of the model~\cite{zi1997comparison,zhu2005size,grujic2009use}. This ambiguity causes difficulties in the numerical implementation of PCM. Yet another ambiguity comes from the use of an arbitrary one-dimensional dispersion curve, the steepness of which is different in different works~\cite{ager1991spatially,yoshikawa1995raman,chaigneau2012laser}. Although these problems could be solved by coupling PCM with quantum-chemical dispersion~\cite{korepanov2020localized}, there is a question of the linewidth: even for a single nanoparticle, the phonon confinement model produces a smooth and broad Raman spectrum. Typically, PCM implies a confinement-induced broadening that links the phonon lifetime with the size~\cite{ager1991spatially}. Additional broadening comes from other mechanisms (broad nanoparticle size distribution or lattice defects). Therefore, the overall linewidth has multiple contributions that are difficult to identify and separate. Nevertheless, the spectrum of a single nanodiamond would have a comb-like shape with multiple narrow peaks~\cite{jenkins1980raman,zhang2005signature,filik2006ramanAbinitio,filik2006ramanDiamondoids,li2010convergence} as, e.g., that of fullerenes~\cite{kimbrell2014analysis}. This comb-like single-particle spectrum shape can be obtained microscopically both using the dynamical matrix method~\cite{koniakhin2018raman} and ab-initio approaches~\cite{li2010convergence}.
	
	Experimentally, published spectra and size estimates vary widely for similar nanodiamond samples~\cite{shenderova2011nitrogen,koniakhin2015molecular,stehlik2016high,kurkin2025detonation}. Various techniques, e.g., centrifugation and fractionation, coupled with transmission electron microscopy (TEM), X-ray diffraction (XRD), and X-ray photoelectron spectroscopy (XPS), indicate that surface chemistry, aggregation, and point-like defects could bias Raman-only size analysis based on simple PCM~\cite{ozerin2008x,mermoux2014surface,koniakhin2015molecular,koniakhin2018ultracentrifugation,trofimuk2018effective}. Modified PCM schemes have been proposed, incorporating anisotropic dispersion~\cite{osswald2007phonon,korepanov2014communication,korepanov2017carbon},  and improved weighting kernels~\cite{zi1997comparison}. At the same time, separating size quantization, disorder, strain, and heating contributions~\cite{chaigneau2012laser,vlk2022nanodiamond} remains ambiguous. 
	
	The usage of the microscopic dynamical matrix method for diamond nanoparticle phonon modes, followed by the bond polarization model for Raman spectra calculation for these modes, solves a significant part of the PCM issues~\cite{koniakhin2018raman}. The dispersion (and, further, the discrete nanoparticle phonon mode energies) is derived based on the Keating model with well-documented elastic coefficients for diamond and other materials (Si and Ge). This approach gives the Raman spectrum shape with discrete peaks, qualitatively reproducing the first-principles calculations~\cite{filik2006ramanAbinitio,koniakhin2018raman}. As a result, the ambiguities in the $\alpha$ parameter and phonon dispersion vanish. Moreover, this microscopic approach was reformulated as a scalar elasticity-like theory for optical phonons and the continuous version of the bond polarization model~\cite{utesov2018raman} to derive Raman scattering intensities (form factors). The subsequent theoretical studies based on Green's function approach, supplied with the dynamical matrix calculations, emphasize the co-existence and competition of confinement effect and lattice disorder (including point defects and substitutional impurities) in setting the Raman peak~\cite{utesov2020lifetimes,koniakhin2020lifetimes}. Refined modeling shows that impurity-induced scattering can produce shifts and broadening comparable to the confinement, and neglecting such effects biases size estimates~\cite{koniakhin2024raman, osswald2007phonon}.
		
	Nevertheless, the community lacks a broadly accessible, standardized, unambiguous, and user-friendly numerical tool for nanodiamond Raman spectra calculation and fitting, embodying more reliable models beyond PCM. Most published frameworks are embedded in specialized simulation packages or available only as fragmented supplementary codes, which limits their adoption. Moreover, inversion of Raman spectra into reliable size distributions requires a stable forward model and fitting algorithms with uncertainty diagnostics -- features rarely packaged together.
	
	In this work, we present an open, documented numerical toolbox that implements a physically motivated forward model for theoretical nanodiamond Raman spectra simulations and experimental spectra treatment with the sample features extraction. The underlying theory combines the size quantization effect for phonon modes in the continuous approach~\cite{utesov2018raman}, realistic phonon dispersion obtained via the microscopic Keating model~\cite{koniakhin2018raman}, and accounts for the explicit impurity/disorder contributions~\cite{koniakhin2024raman}. This theory solves the forward problem, i.e., derives Raman spectra from the predefined nanodiamond size distribution and defect parameters. The resulting spectrum-fitting tool includes two complementary inversion strategies -- a neural network-based regression trained on the simulated spectra and a Metropolis-style stochastic refinement -- to extract particle size distributions and disorder parameters from experimental powder spectra. In the neural network (NN) approach, the Raman spectrum $\mathbf{X}$ is mapped to the size distribution histogram and defect parameters encoded in vector $\mathbf{Y}$. So, the neural network is trained to effectively solve the inverse problem of taking the spectrum and yielding the nanopowder parameters. The trained neural network plays here the role of an advanced multi-dimensional function from the Raman count domain to the size histogram and defect parameter domain. This function absorbs all underlying physics of optical phonon size quantization and lattice defect effects and provides fast spectrum treatment. Representation of this function as a neural network is practically useful due to modern progress in the area. The complementary Metropolis approach starts with the random size distribution and defect parameters and gradually updates them to achieve a minimal deviation between the calculated Raman spectrum and the target (experimental) one. Finally, for the neural network and Metropolis approaches, we demonstrate the effectiveness of treatment for both synthetic spectra and experimental data.
	
	The developed tool is freely available via the following ways:
    \begin{itemize}
        \item Online tool \href{https://nanoraman.pythonanywhere.com/}{nanoraman.pythonanywhere.com/}
        \item The source code ready for the standalone run \href{https://github.com/KoniakhinSV/Nanoparticle\_Raman}{(github.com/KoniakhinSV/Nanoparticle\_Raman)}.
        \item The \href{https://arxiv.org}{arXiv} version of the paper will be updated to contain \textit{actual links} to the script source and to the website.
    \end{itemize}
    The actual links to the online version and to the GitHub repository will be updated and kept active in the arXiv version of the paper. The README file on the GitHub repository will contain a link to the actual version of the site as well.

	%We show that (i) including impurity/disorder terms is essential to avoid systematic size bias; (ii) model selection and parameter uncertainty must be explicitly reported; and (iii) a well-documented, user-friendly implementation lowers the barrier for experimentalists to adopt physics-based Raman analysis and reliably obtain quantitative size and disorder metrics from their data.

	%===============================================

	%===============================================

	%===============================================

	%===============================================

	%===============================================

	\section{Methods}

%>>>>>>>>>>>>>>>>>>>>>>>>>>>>>>>>>>>>>>>>>>>>>>>>>>>>>>>>>>>>>>>>>>>>>>>>>>>>>>>>>>>>>>>>>>>>>>>
\begin{figure*}[htbp]
		\centering
		\includegraphics[width=0.4\linewidth]{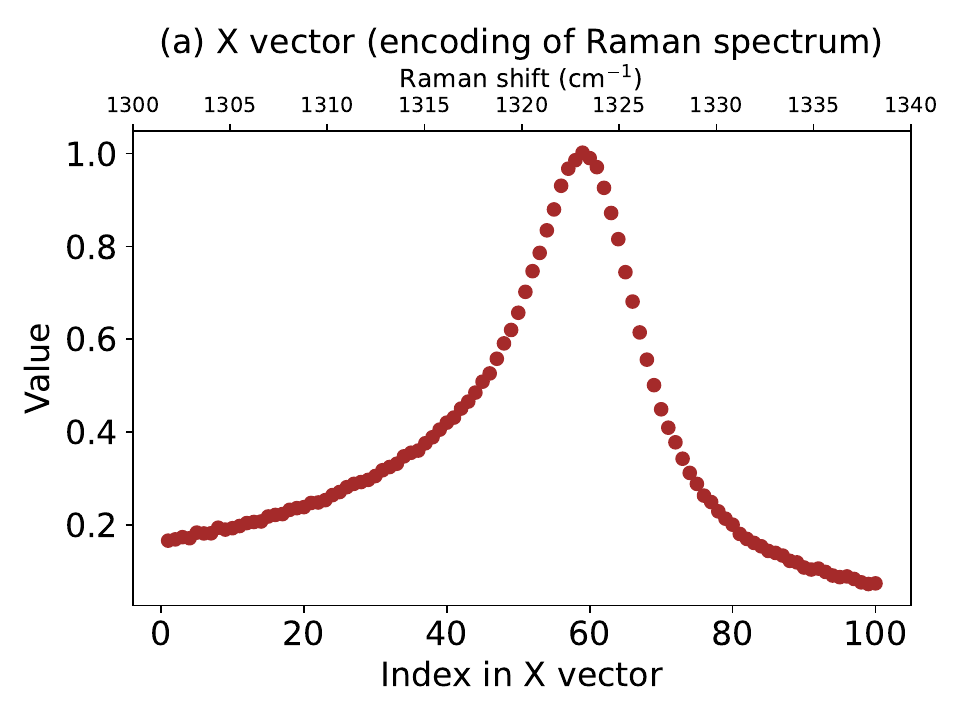}\includegraphics[width=0.6\linewidth]{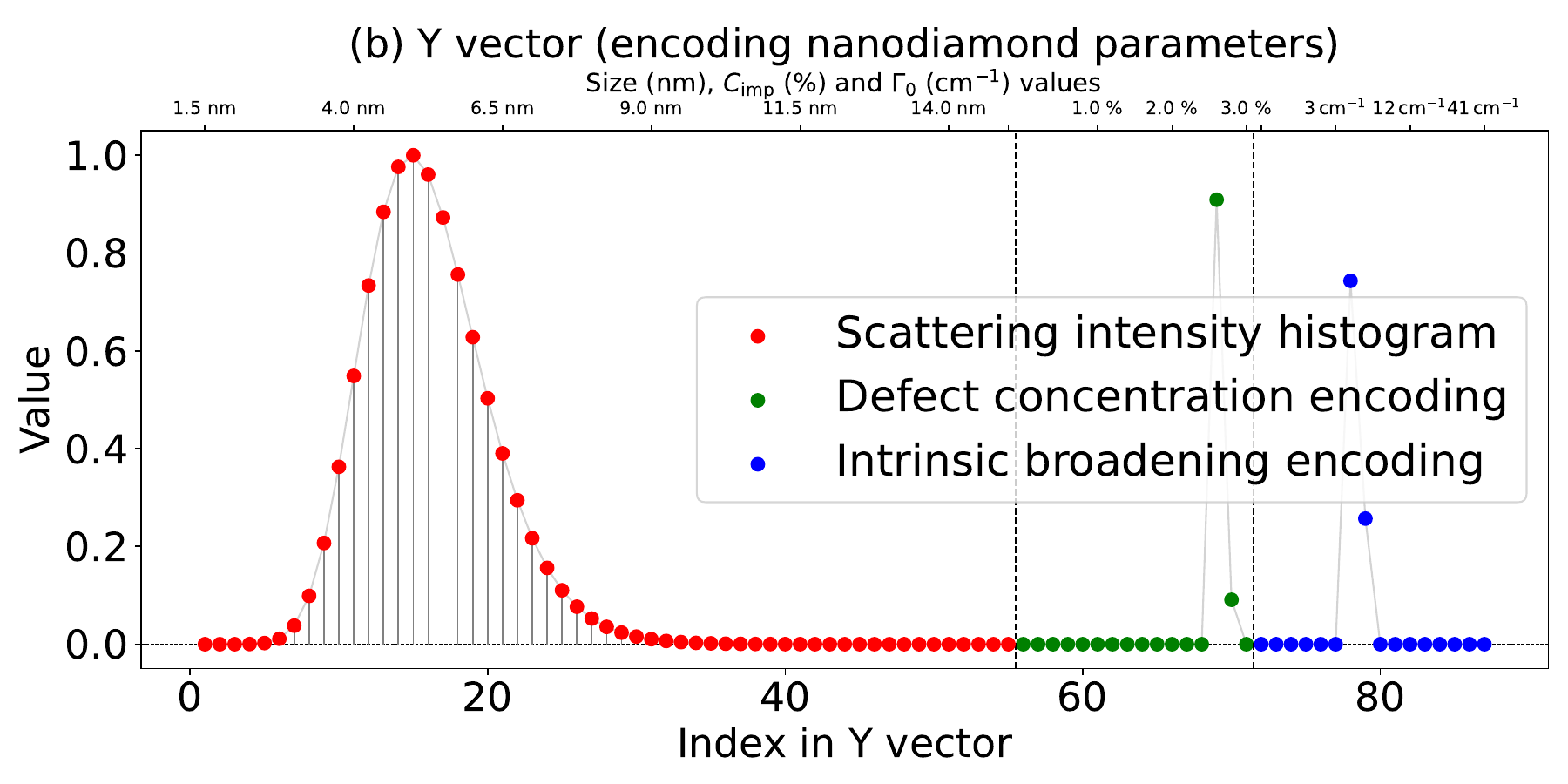}
		\caption{Panel (a) shows Raman spectrum vector $\mathbf{X}$ derived using continuous bond polarization model and disorder theory for nanodiamond Raman spectra based on the size distribution and defect parameters encoded in $\mathbf{Y}$. (b) Nanoparticle size histogram is encoded in the first 55 bins (red dots). The disorder concentration is $C_{\rm imp} \approx 2.8$ (encoded as 0.2 times the largest element in green dot one-hot vector) and the intrinsic broadening $\Gamma_0 = 4.4$~cm$^{-1}$ (encoded as 1.28 to the power of the largest element in blue dot one-hot vector).}
		\label{fig1_encoding}
\end{figure*}
%>>>>>>>>>>>>>>>>>>>>>>>>>>>>>>>>>>>>>>>>>>>>>>>>>>>>>>>>>>>>>>>>>>>>>>>>>>>>>>>>>>>>>>>>>>>>>>>

	\subsection{Theoretical Framework for Raman Spectra of Nanodiamonds}
	The Raman spectrum of a single diamond nanoparticle has a discrete structure with the signals from various phonon modes characterized by specific quantum numbers $n$. So, the Raman spectrum for a diamond nanoparticle of size (diameter) $L = 2R$ can be derived as a sum of Lorentzian peaks centered at the respective vibrational mode energies $\omega_n$ and broadened with the mechanisms described below: 
	\begin{equation}
		I(\omega; R) = \sum_{n=1}^{n_{\text{max}}} I_n \cdot \frac{(\Gamma_n/2)^2}{[\omega - \omega_n(R)]^2 + (\Gamma_n/2)^2}.
	\end{equation}
	In the equation above, $I_n$ is the scattering intensity of a given mode. In the nanoparticle of arbitrary geometry, optical phonon modes behave like standing waves in the resonator. Their spatial profile  obeys the Helmholtz equation $\Delta Y + q^2 Y=0,  \qquad Y|_{\partial \Omega} = 0$ with plane wave dispersion $\omega^2=C_2-C_1 q^2$, see Eqs.~(22)-(24) of Ref.~\cite{utesov2018raman}. As it was previously demonstrated, the spectra for various nanoparticle shapes are qualitatively similar. This picture is at least correct for the current experimental state-of-the-art, i.e., measurement precision, background subtraction-induced ambiguity, and inaccessibility of measuring the single-particle spectrum. For the model case of a spherical particle, $n$ absorbs principal, azimuthal, and magnetic quantum numbers (in the hydrogen atom notation). Within the continuous form of the bond polarization model, the scattering intensity $I_n~=~|\int Y_n(\mathbf{r})d\mathbf{r}|^2$, which resembles the form factor definition in diffraction. Due to the symmetry, only the modes with zero azimuthal quantum number contribute, and the corresponding scattering form factors read
	\begin{equation}
		I_n = 6 \cdot \dfrac{4}{3} \pi R^3 / (\pi n)^2.
	\end{equation}
	By spanning over multiple modes, we obtain the comb-like shape of the spectrum. However, with the increase of mode number $n$, Raman intensity rapidly decreases. For the nanoparticles of realistic shapes (e.g., truncated octahedra and similar), one observes symmetry breaking and level splitting, as well as in the disordered nanoparticles. It will lead to an increase in the number of peaks in the spectrum, as it was shown in Ref.~\cite{li2010convergence}. At the same time, the atomistic structure of nanoparticles does not affect their vibrational properties until very small atomic clusters are addressed~\cite{combe2009acoustic}. 
	
	Under the assumption of the parabolic dispersion, the mode position is written as
	\begin{equation}
		\omega_n = A - B \dfrac{q_n^2 a_0^2}{8} + \delta\omega,
	\end{equation}
	where effective wave vector $q_n = \pi n / R$ and $\delta\omega$ is the disorder-induced peak position shift. We employ the dispersion parameters $A = 1333$ cm$^{-1}$, $B = 85$ cm$^{-1}$, and $a_0 = 0.357 \times 10^{-7}$~cm. The broadening is a combination of intrinsic size-independent linewidth $\Gamma_0$ and disorder-induced inhomogeneous broadening. The former is a homogeneous broadening combining the temperature effects and spectrometer linewidth. The disorder-induced broadening causes effective averaging of the Raman signal from the nanoparticles with various concentrations and configurations of lattice defects.
	\begin{equation}
		\Gamma = \Gamma_0 + \Gamma_{\text{dis}}
	\end{equation}

	As shown in Ref.~\cite{koniakhin2024raman}, the isotope disorder in masses or introducing heavy substitution impurities do not lead to a significant effect on Raman spectra, and the only valuable type of lattice defects are light impurities, vacancies, or complex defects containing a vacancy (e.g., NV-centers). In that case, the following Raman peak shift $\delta\omega$ and inhomogeneous broadening $\Gamma_{\text{dis}}$ can be obtained:
	\begin{align}
		\delta\omega &= -2.6~\mathrm{cm}^{-1} \cdot C_{\text{imp}}, \\
		\Gamma_{\text{dis}} &= 1.5~\mathrm{cm}^{-1} \cdot C_{\text{imp}} \cdot \frac{3.0~\mathrm{nm}}{L},
	\end{align}
	where $c_{\text{imp}}$ is the impurity concentration in terms of \textit{per cent} and $L = 2R$ is the particle size in nanometers. For a distribution of sizes $\{ L_i \}$ with normalized counts $N_i$, the total spectrum is:
	\begin{equation}
		I_{\text{total}}(\omega) = \frac{1}{\max(I_{\text{total}})} \sum_i N_i \cdot I(\omega; R_i).
	\end{equation}

	For powder spectra calculations, one can use several standard distribution functions. Along with the standard normal distribution, we use the natural distribution for nanodiamonds, namely the log-normal one~\cite{maul2005lognormal,kovavrik2020particle}:
	\begin{equation}
        \label{eq_PDF}
		\mathrm{PDF}(L,\mu,\sigma) = \frac{1}{L \sigma \sqrt{2\pi}} \exp\left(-\frac{(\ln L - \ln \mu)^2}{2\sigma^2}\right),
	\end{equation}
    which, upon discretization, gives us the required $N_i$'s.

	\subsection{Neural Network Approach for Spectrum Fitting}

%>>>>>>>>>>>>>>>>>>>>>>>>>>>>>>>>>>>>>>>>>>>>>>>>>>>>>>>>>>>>>>>>>>>>>>>>>>>>>>>>>>>>>>>>>>>>>>>	
	\begin{figure}[h]
		\centering
		\includegraphics[width=0.9\linewidth]{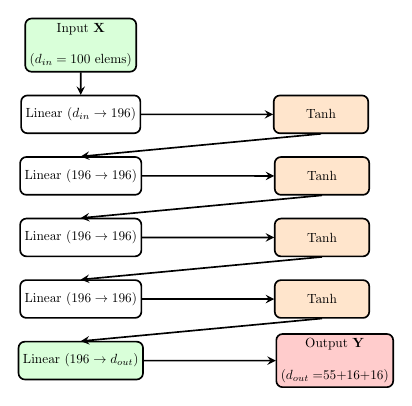}
		\caption{Architecture of the neural network that was shown to efficiently encode the multidimensional function which maps from the Raman spectrum to the corresponding nanodiamond size distribution histogram and defect parameters.}
		\label{fig:ramannet}
	\end{figure}
%>>>>>>>>>>>>>>>>>>>>>>>>>>>>>>>>>>>>>>>>>>>>>>>>>>>>>>>>>>>>>>>>>>>>>>>>>>>>>>>>>>>>>>>>>>>>>>>	

	\textit{Training dataset creation.} The theory described above allows Raman spectra calculation from the given nanoparticle size distribution and disorder parameters. It can be used for the creation of a synthetic training dataset for the inverse problem, i.e., size distribution reconstruction from the available Raman spectrum. This dataset has the following structure.
	The size distribution (by number of particles) has the meaning of a histogram with 55 bins centered at $L_i$ and covering the range from 1.5~nm to 15~nm. To generate it, we sum from one to three log-normal and normal distributions, defined by the most probable size and the standard deviation. The normal distribution has a standard shape, and the log-normal one reads as [cf. Eq.~\eqref{eq_PDF}]
	\begin{equation}
		N_i = \sum_{\alpha=1}^3\mathrm{PDF}(L_i,\mu_{\alpha},\sigma_{\alpha})
	\end{equation}
	where $\mu_\alpha \in [2.0, 8.0]$~nm and $\sigma_\alpha \in [0.1, 0.9]$~nm give characteristic size and broadening in $\alpha$-th peak of the multimode size distribution. Their values are taken randomly from the uniform distributions of the corresponding range. Normalization is performed to the maximal value in the histogram. As a result, one has a 55 element vector $\mathbf{N} = \{ N_1,\dots,N_{55}\}$ for nanodiamond size distribution representation, see Fig.~\ref{fig1_encoding}.

	The impurity concentration (in terms of percent) was discretized into 16 values as $C_{\text{imp}} = 0.2 (k-1)$~cm$^{-1}$, which gives a 0.0--3.0~cm$^{-1}$ range in steps of 0.2~cm$^{-1}$. This range covers the reported values of concentrations in detonation synthesis nanodiamonds~\cite{chang2016counting}, known as the most defect type of diamond nanoparticles. For creating the dataset sample, the random choice of $C_{\text{imp}}$ was performed. In the training dataset, it was encoded as a proportional one-hot vector. In the case of strict coincidence of $C_{\text{imp}}$ with discrete values, one has $\mathbf{c} = \{0, 0, \dots , \underbrace{1}_{k-\mathrm{th}}, \dots,0 \}$. The intrinsic and instrumental broadening combination was discretized using the logarithmic scale as $\Gamma_0 = 1.28^{l-1}$ with $l=1,\dots,16$, which yields 1.3--40.5~cm$^{-1}$ range. The corresponding one-hot vector is $\mathbf{g} = \{ 0, 0, \dots , \underbrace{1}_{l-\mathrm{th}}, \dots,0 \}$.	
	
	The output vector $\mathbf{Y}$ used for the training is a concatenation of the size distribution histogram, impurity concentration, and broadening encoding vectors: $\mathbf{Y} = \{\mathbf{N},\mathbf{c},\mathbf{g} \}$ or in the explicit form:
	\begin{equation}
		\mathbf{Y} = \left[ \underbrace{N_1, \dots, N_{55}}_{\text{size counts}}, \underbrace{0,0,\dots,1,\dots,0}_{C_{\rm imp}}, \underbrace{0,0,\dots,1,\dots,0}_{\Gamma_0}\right],
	\end{equation}
	Its length is 87 elements: 55 for size counts + 16 for $C_{\text{imp}}$ + 16 for $\Gamma_0$.

	Based on the size distribution histogram and defect parameters encoded in $\mathbf{Y}$, the continuous bond polarization model allows us to calculate the Raman spectrum $\mathbf{X}$. For each sample, 1\% of noise was added for stability. The resulting spectrum is stored as intensity counts $\mathbf{X}_j = I_{\text{total}}(\omega_j)$ for the 100 equidistant Raman shift values $\omega_j$ in the range from 1300~cm$^{-1}$ to 1340~cm$^{-1}$. The training dataset for the inverse problem, i.e.,  size distribution prediction based on Raman spectrum, contained $4\cdot10^5$ pairs $(\mathbf{X} \rightarrow \mathbf{Y})$. Fig.~\ref{fig1_encoding} shows the example of the spectrum vector $\mathbf{X}$ and nanoparticle powder parameter vector $\mathbf{Y}$.

	\textit{Network Architecture.} A fully connected feedforward neural network with 4 hidden layers (196 neurons each) and $\tanh$ activation was implemented, see Fig.~\ref{fig:ramannet}. In accordance with the Raman spectrum encoding vector and the size distribution encoding vector, the input and output dimensions were 100 and 87 = 55 + 16 + 16, respectively. The architecture comprises:
	\begin{itemize}
		\item \textbf{Input layer}: 100 inputs (Raman spectrum representation)
		\item \textbf{Hidden layers}: 4 layers of 196 neurons with tanh activation
		\item \textbf{Output layer}: 87 outputs (55 for size counts + 16 for $C_{\text{imp}}$ + 16 for $\Gamma_0$)
	\end{itemize}
	The model was trained to minimize the mean square error (see below) between predicted and true $\mathbf{Y}$ using the synthetic data described above. After the training, the size distribution was directly recovered from the first part of the vector $\mathbf{Y}$. The respective $\mathbf{Y}$ parts encode the resulting broadening and defect concentration.
	
	The chosen model architecture, including the hidden layer dimension and activation function, is not the only possible choice. The total parameter number is sufficient to represent a complex function from the Raman spectrum domain to the size distribution histogram and disorder parameter domain, absorbing the underlying physical model. At the same time, no overfitting takes place. To refine the NN predictions, the Raman spectra were calculated for 5 impurity concentrations encoded by 5 maximal elements of the $\mathbf{c}$ vector and for 5 size-independent broadenings $\Gamma_0$ encoded by 5 maximal elements of the vector $\mathbf{g}$. As a final result, the spectrum with minimal mean square deviation with respect to the experimental one was chosen.

	%===========================================

	\subsection{Metropolis Optimization for Spectrum Fitting}
    
\begin{figure}[htbp]
	\centering
	\includegraphics[width=0.99\linewidth]{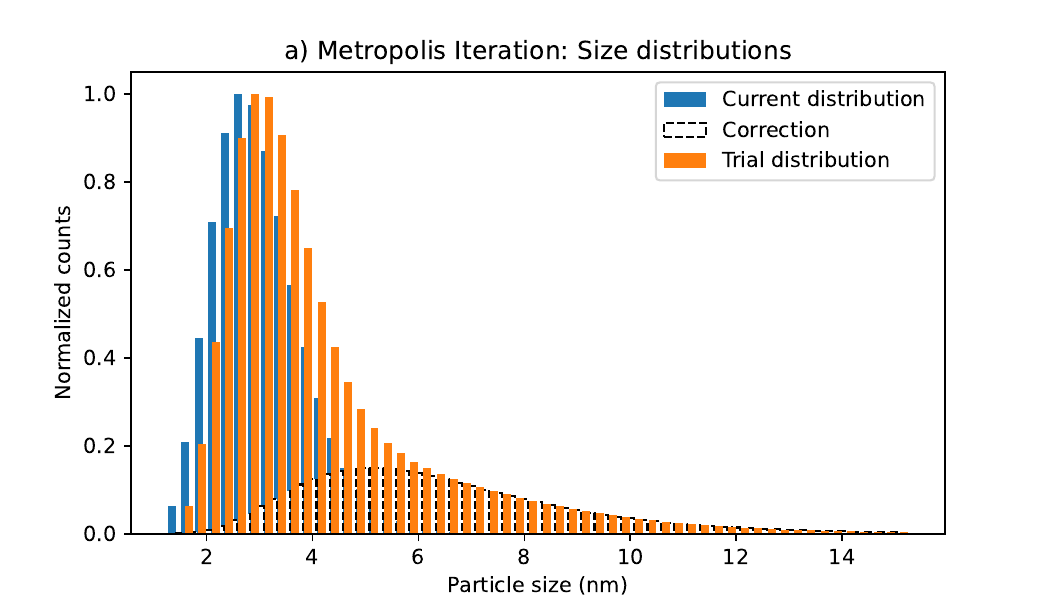}\\
    \includegraphics[width=0.99\linewidth]{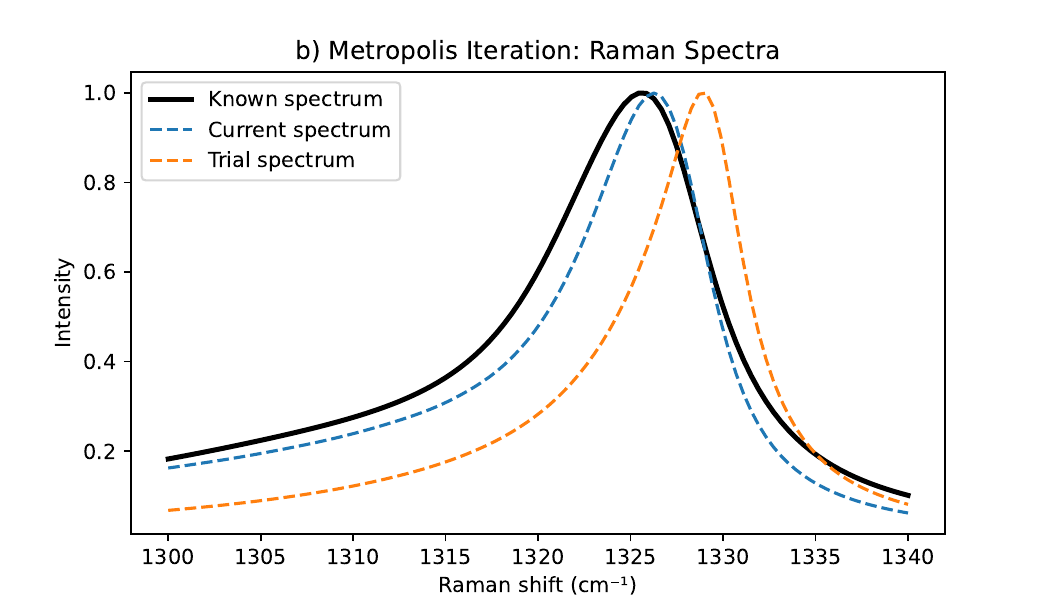}
	\caption{Schematics of Metropolis spectrum fitting procedure. Panel a) shows the current size distribution updated by a correction with the trial distribution as a result. Panel b) shows the spectra simulated for the current size distribution (and also $c_\text{imp}$ and $\Gamma$ parameters) and for the trial one. The trial correction is declined by the algorithm because it results in a worse spectrum fit.} 
	\label{fig_metropolis_scheme}
\end{figure}

	The Metropolis algorithm minimizes the mean squared deviation (MSD) between experimental ($I_{\text{exp}}$) and reconstructed ($I_{\text{rec}}$) spectra:
	\begin{equation}
		\text{MSD} = \frac{1}{N} \sum_{i=1}^{N} \left[ I_{\text{exp}}(\omega_i) - I_{\text{rec}}(\omega_i) \right]^2.
	\end{equation}
	Starting from a random log-normal size distribution, impurity concentration $c_{\text{imp}}^{(0)}$, and broadening $\Gamma_0^{(0)}$, the algorithm iteratively updates the defect parameters and size distribution:
    \begin{equation}
    \begin{aligned}
        \mathbf{N}_{\text{trial}} &= \mathbf{N}_{\text{current}} + \eta \cdot \mathbf{N}_{\text{corr}},\quad \eta \in [-0.05, 0.05] \\
        C_{\text{imp}}^{\text{trial}} &= C_{\text{imp}}^{\text{current}} + \delta C, \quad \delta c \in [-0.01, 0.01], \\
        \Gamma_0^{\text{trial}} &= \Gamma_0^{\text{current}} + \delta \Gamma, \quad \delta \Gamma \in [-0.05, 0.05] \, \text{cm}^{-1},
    \end{aligned}
    \end{equation}
	where $\mathbf{N}_{\text{corr}}$ is the distribution given by Eq.~\eqref{eq_PDF} with the random centroid and broadening. The $\eta$ parameter is also taken randomly from the outlined range, as well as the defect concentration and intrinsic broadening corrections. If trial parameters yield better MSD than the current parameters, then the trial ones are accepted, and the next iteration takes place. This scheme corresponds to the Metropolis algorithm with zero temperature. Fig.~\ref{fig_metropolis_scheme} illustrates one step of the Metropolis approach. It shows the changes in size distribution (panel a) and respective Raman spectra (panel b) for a single update step~($\eta$ parameter is taken 10 times larger for better graphical distinguishability). Note that in the software, there is a possibility to fix the impurity concentration and/or size-independent broadening for the Metropolis fit to narrow the parameter search space.
	
%>>>>>>>>>>>>>>>>>>>>>>>>>>>>>>>>>>>>>>>>>>>>>>>>>>>>>>>>>>>>>>>>>>>>>>>>>>>>>>>>>>>>>>>>>>>>>>>		
%>>>>>>>>>>>>>>>>>>>>>>>>>>>>>>>>>>>>>>>>>>>>>>>>>>>>>>>>>>>>>>>>>>>>>>>>>>>>>>>>>>>>>>>>>>>>>>>	
    
In summary, both NN and Metropolis approaches have a Raman spectrum as an input and produce the size distribution, $C_{\text{imp}}$, $\Gamma_0$, and the reconstructed spectrum, with the results packaged into ZIP archives containing CSV files and plots.

\subsection{Background subtraction}

We have also implemented the unmanned procedure for the background subtraction refinement. It is based on the neural network approach. The input $\mathbf{X}$ vector is a Raman spectrum with a background. The output $\mathbf{Y}$ vector is a pure Raman spectrum. We generate a $\mathbf{Y}$ vector from a random size distribution and disorder properties as previously described for the neural network approach for the spectra fit. The background to be added is a linear function defined by the random shifts for 1300~cm$^{-1}$ and 1340~cm$^{-1}$. The structure of the neural network is the same, except that the output dimension is the standard Raman-spectrum encoding vector, i.e., 100 elements. The background subtraction procedure demonstrates perfect precision for the synthetic data. All versions of the package allow for both the application of the background subtraction and the treatment of the original spectra without the subtraction. Thus, the additional functionality of our package is the refinement of the background subtraction procedure.

%==========================================================	

%==========================================================	

%==========================================================	

%==========================================================	
\section{Results}

To characterize the size distribution histogram, the useful value is the typical nanoparticle size with respect to the number of particles
\begin{equation}
	\widetilde{L}_N = \frac{\sum_{i=1}^N P(L_i) L_i}{\sum_{i=1}^N P(L_i)}
\end{equation}
and with respect to volume fractions:
\begin{equation}
	\widetilde{L}_V= \frac{\sum_{i=1}^N P(L_i) L_i^4}{\sum_{i=1}^N P(L_i) L_i^3}
\end{equation}
Noteworthy, since the Raman intensity is proportional to the particle volume, the latter quantity indicates the particles providing the maximum in the Raman spectrum. For instance, in the model case of the bimodal size distribution with equivalent number of 2~nm and 10~nm particles, the first particle-number definition will yield $\widetilde{L}_N =6 $~nm while the second volume-based definition $\widetilde{L}_V \approx 9.9$~nm. The bigger particles evidently dominate in the Raman spectrum, SAXS, and XRD, and in that case $\widetilde{L}_V$ is a better metric to compare with. At the same time, HRTEM and AFM data can be better compared with $\widetilde{L}_N$.

\begin{figure*}[htbp]
		\centering
\includegraphics[width=0.49\linewidth]{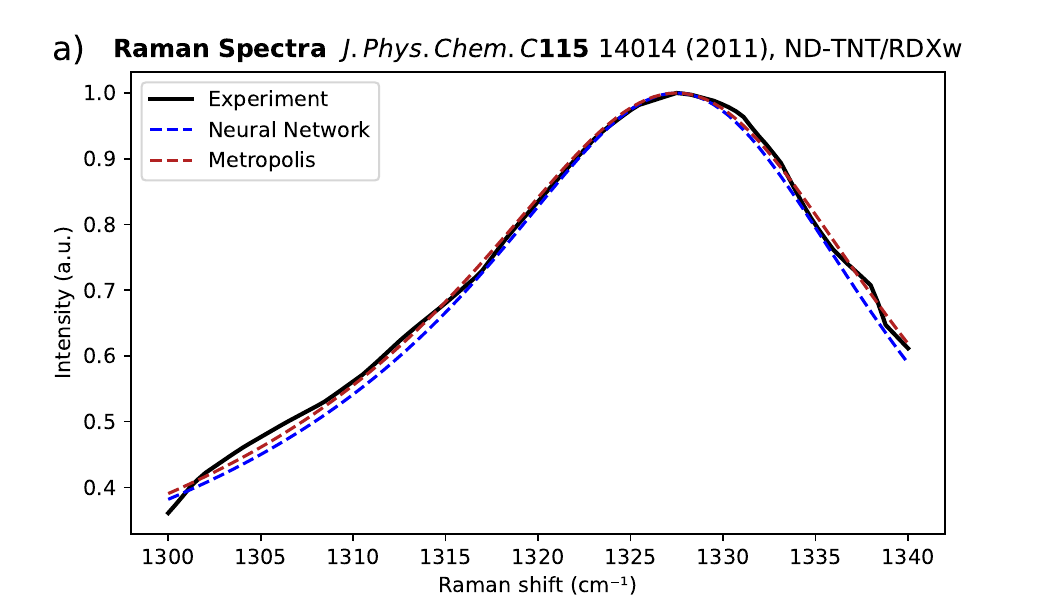}\includegraphics[width=0.49\linewidth]{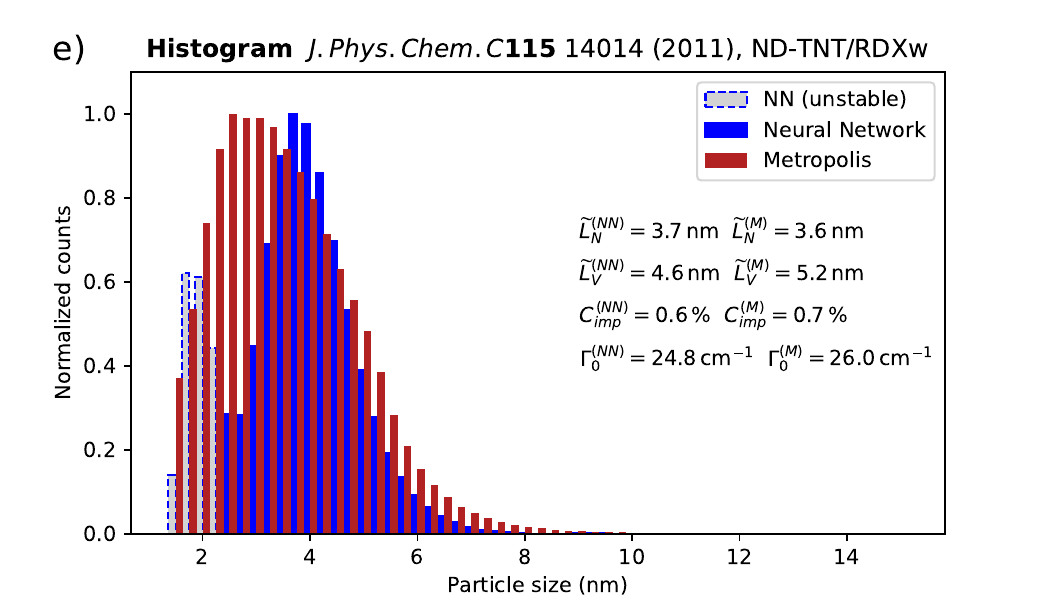} \\ \includegraphics[width=0.49\linewidth]{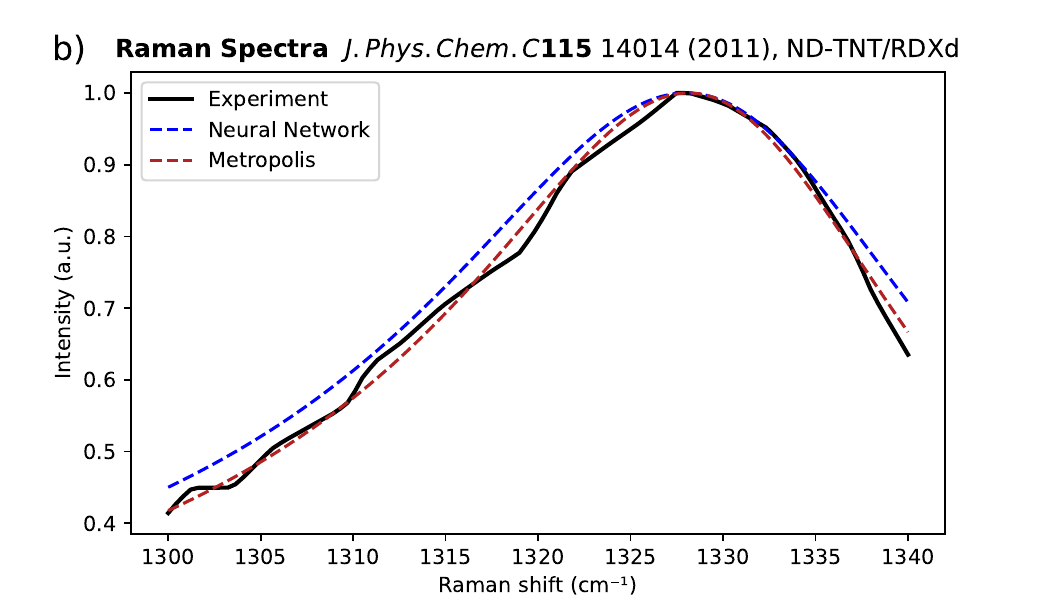}\includegraphics[width=0.49\linewidth]{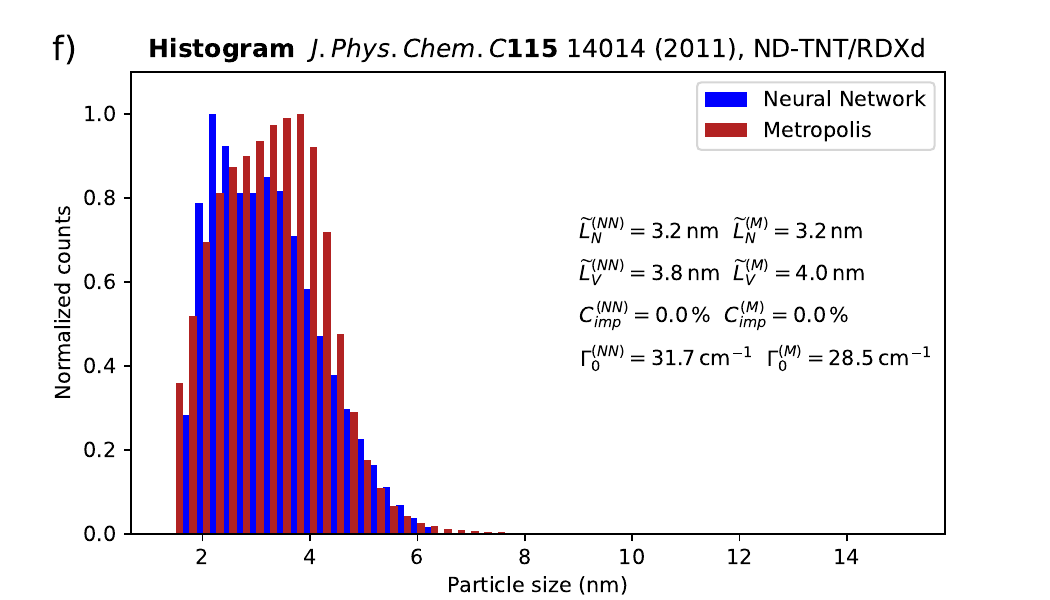} \\ \includegraphics[width=0.49\linewidth]{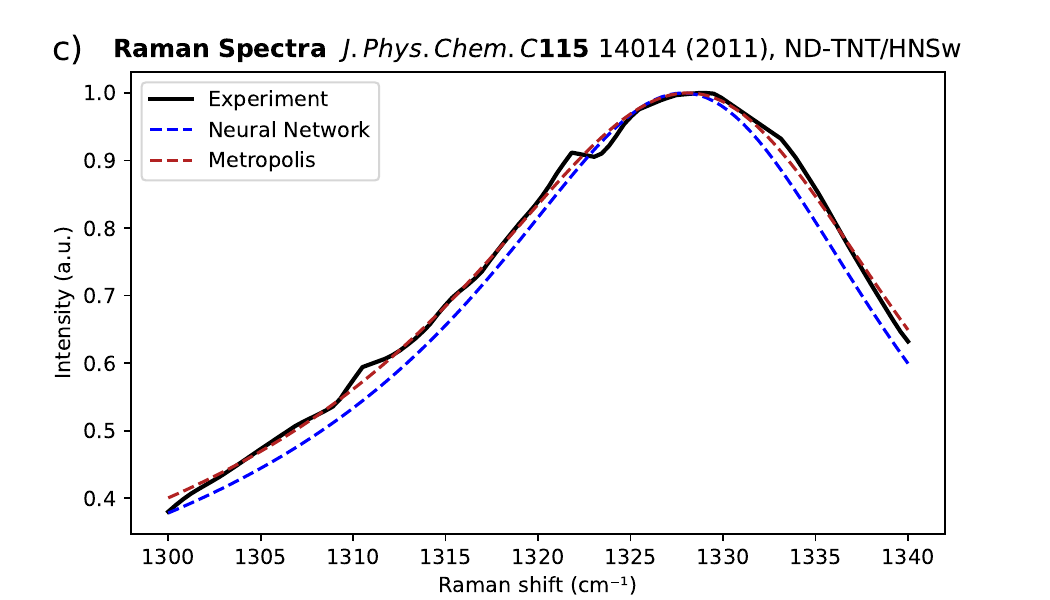}\includegraphics[width=0.49\linewidth]{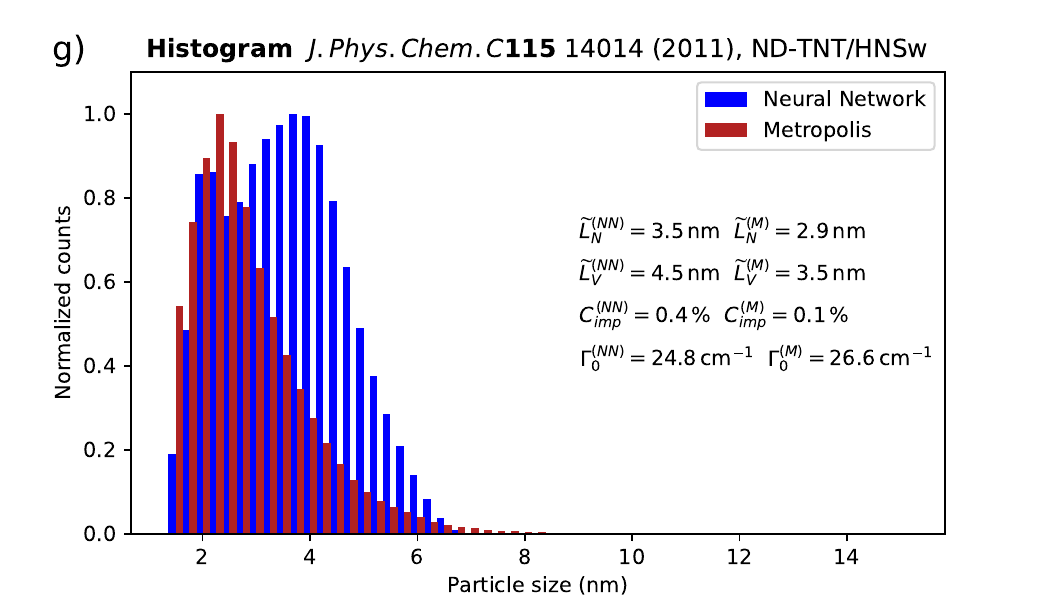} \\ \includegraphics[width=0.49\linewidth]{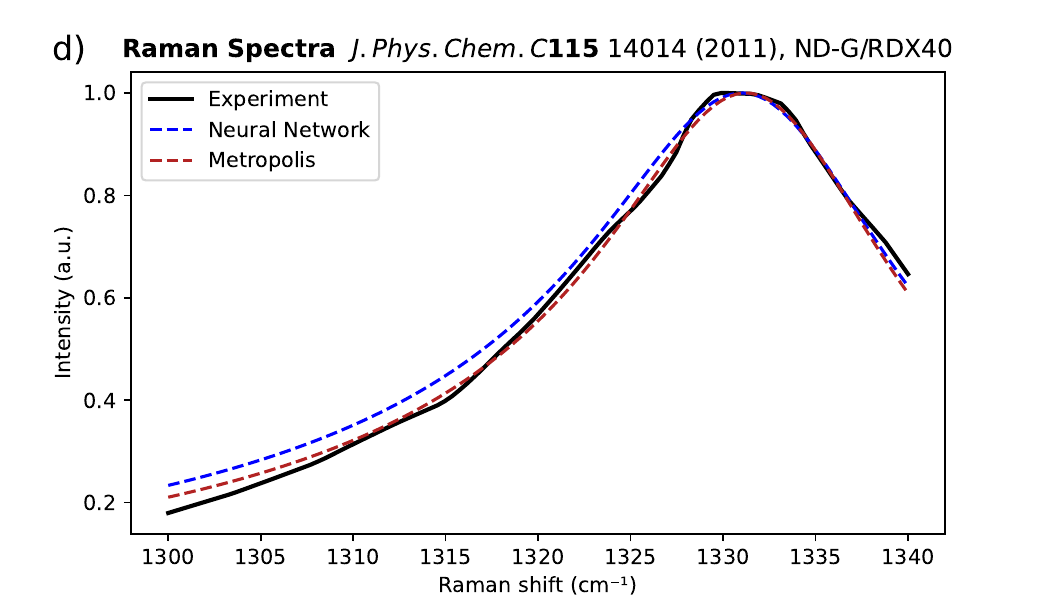}\includegraphics[width=0.49\linewidth]{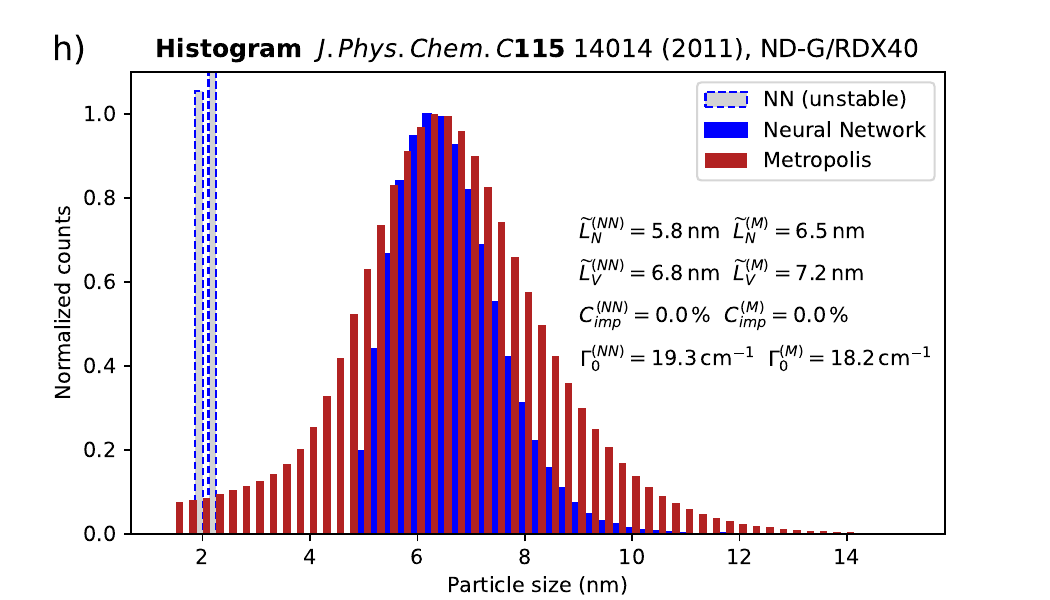} \\ 
		\caption{Result of the fits for nanodiamond samples reported in Ref.~\cite{shenderova2011nitrogen}. Panels a),b),c) are for the Raman spectra, and panels e),f),g) are for the reconstructed size distributions of detonation synthesis nanodiamond samples ND-TNT/RDXw, ND-TNT/RDXd, and ND-TNT/HNSw of typical size 4-5~nm, correspondingly. Histograms are given by the number of particles.  Panels d) and h) show the Raman spectrum and the histogram of the sample ND-G/RDX40 obtained from a graphite/RDX mixture and having a size of 9~nm. For spectra (panels a-d), the black curves are the spectra, and the dotted green curves are the original spectra before background subtraction refinement. The blue dashed curves are the Raman spectra reconstructed from the neural network prediction, and the red dashed curves are for those obtained from the Metropolis approach. In the histograms (panels e-h), red bars give the predictions for the Metropolis (M) approach, and blue ones are for the neural network (NN) approach. Gray bars with dashed contours are for the unreliable parts of the neural network-based histograms (see text).}
		\label{fig_exp_shend}
	\end{figure*}

We apply the developed software for the treatment of numerous spectra available in the literature. First, we address the Raman spectra of four different samples presented in Fig.~\ref{fig_exp_shend} from Ref.~\cite{shenderova2011nitrogen} (Shenderova et al., 2011). The details of these samples preparation are given below.

%>>>>>>>>>>>>>>>>>>>>>>>>>>>>>>>>>>>>>>>>>>>>>>>>>>>>>>>>>>>>>>>>>>>>>>>>>>>>>>>>>>>>>>>>>>>>>>>
%>>>>>>>>>>>>>>>>>>>>>>>>>>>>>>>>>>>>>>>>>>>>>>>>>>>>>>>>>>>>>>>>>>>>>>>>>>>>>>>>>>>>>>>>>>>>>>>

Produced by the detonation of a TNT/RDX (40/60) explosive mixture in a wet (ice) cooling medium (wet synthesis) and subsequent extensive chemical purification, the ND-TNT/RDXw sample demonstrates a high nitrogen content of greater than 2.5\% according to elemental analysis.  The ND-TNT/RDXd was produced by detonation of the same explosives using a dry cooling medium and contained 2~wt\% of nitrogen. The ND-TNT/HNSw sample was produced from a TNT/HNS (hexanitrostilbene) explosive mixture using wet cooling. The typical size was the same, and the nitrogen content was 0.5~wt\%. For all of the samples described above, the powder X-ray diffraction method evidences the typical crystallite size of 4~nm. The sample ND-G/RDX40 was produced from a graphite/RDX mixture and demonstrates a crystallite size of around 9~nm. Nitrogen content is low according to elemental analysis.

First of all, Fig.~\ref{fig_exp_shend} demonstrates a high quality of the Raman spectra fit for all the samples.  For the series of 4~nm samples (ND-TNT/RDXw, ND-TNT/RDXd, ND-TNT/HNSw), the predictions based on the Raman spectra yield the characteristic sizes of 4-5~nm. The defect concentration in the units of vacancy-type defect concentration for these samples ranges from 0.0 to 0.7 percent. For the ND-G/RDX40 sample, the size is estimated as 7~nm, and the defect concentration is zero. As one sees, the nanocrystallite size predictions from Raman and XRD measurements correlate and demonstrate coincidence. The lattice defect concentration determined based on Raman spectra cannot be unambiguously interpreted as the content of nitrogen. However, the correlations still exist. The neural network-based procedure of fit can yield a high number of small-sized particles (less than 2.5~nm). This picture takes place because after recalculation to the scattering intensities (proportional to $N_i \cdot L_i^3$), the respective scattering signal is negligible and thus the histogram becomes ambiguous. The region of unstable elements in the histogram has the following features: sizes smaller than 2.5~nm, narrow peaks separated from the main signal, often with a gap between them. For comparison with XRD data, it is natural to choose the typical volume size of nanodiamonds $\widetilde{L}_V$. The automatic background subtraction procedure would provide minor changes to the spectra, which justifies the robustness of the preceding manual procedure.

\begin{figure*}[htbp]
		\centering
    \includegraphics[width=0.49\linewidth]{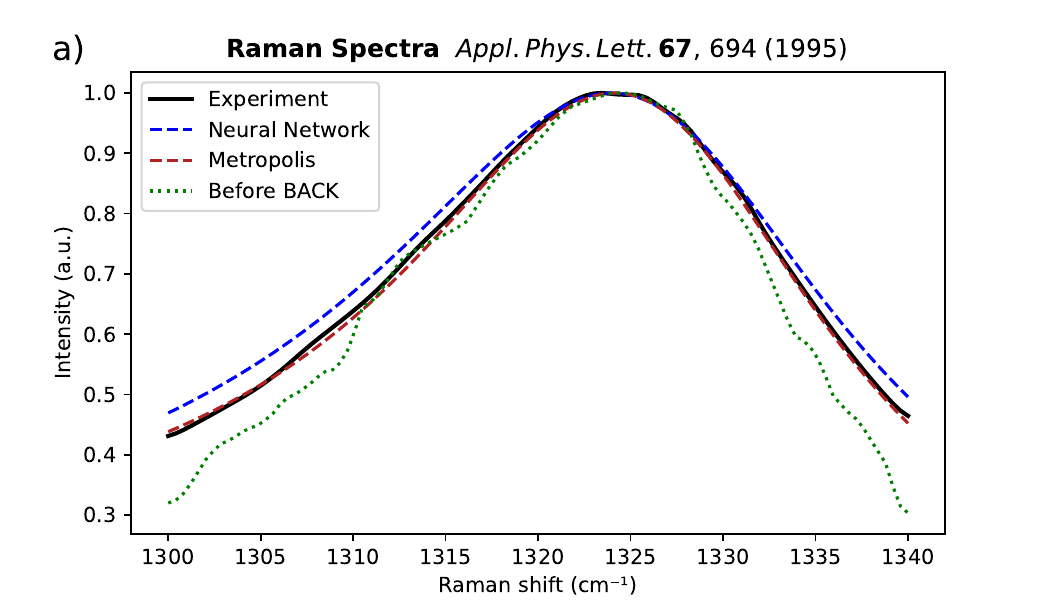}\includegraphics[width=0.49\linewidth]{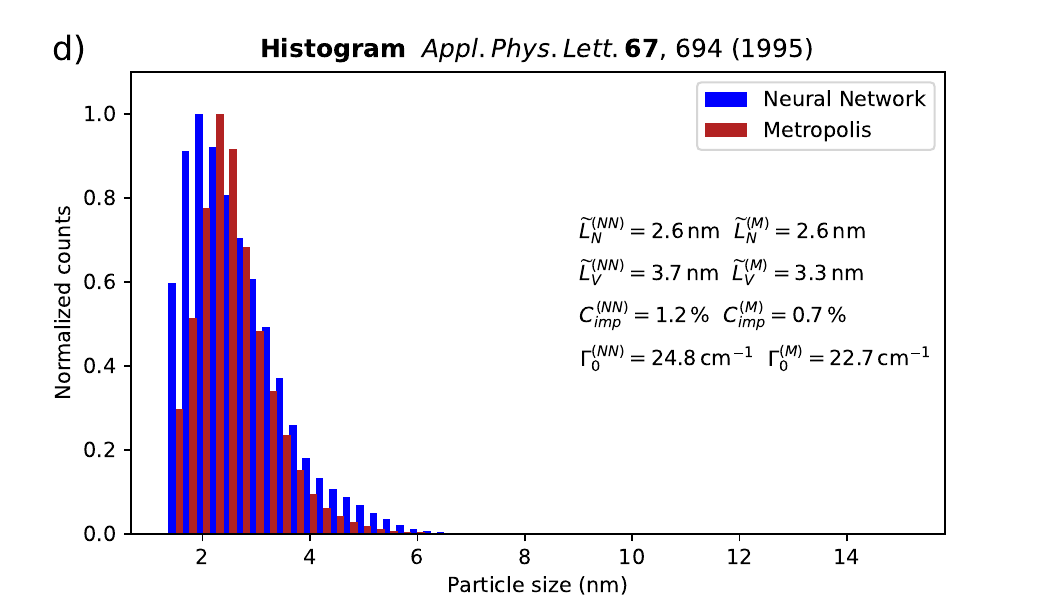} \\ \includegraphics[width=0.49\linewidth]{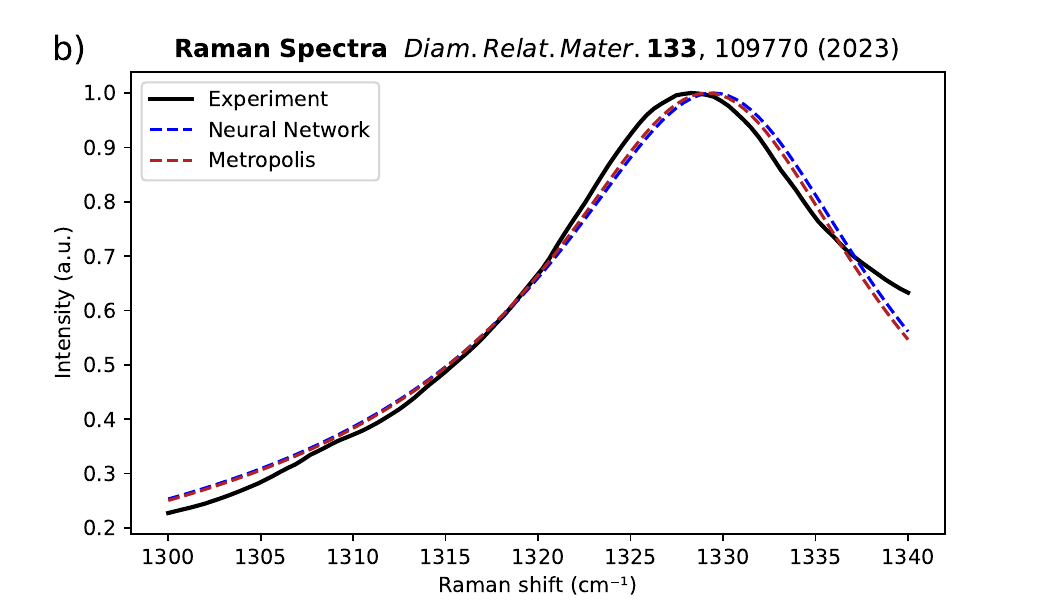}\includegraphics[width=0.49\linewidth]{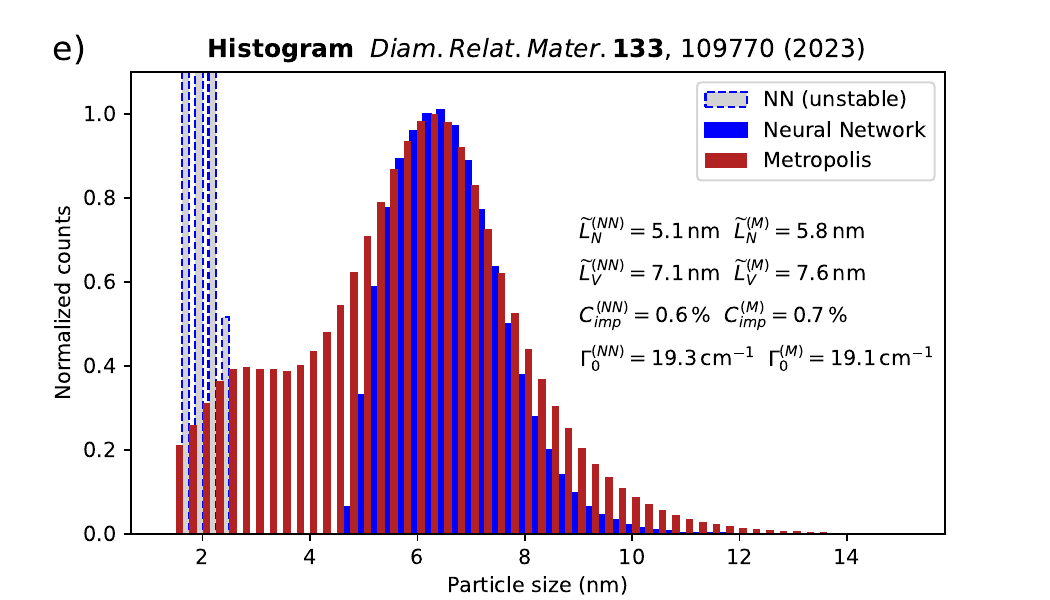} \\ \includegraphics[width=0.49\linewidth]{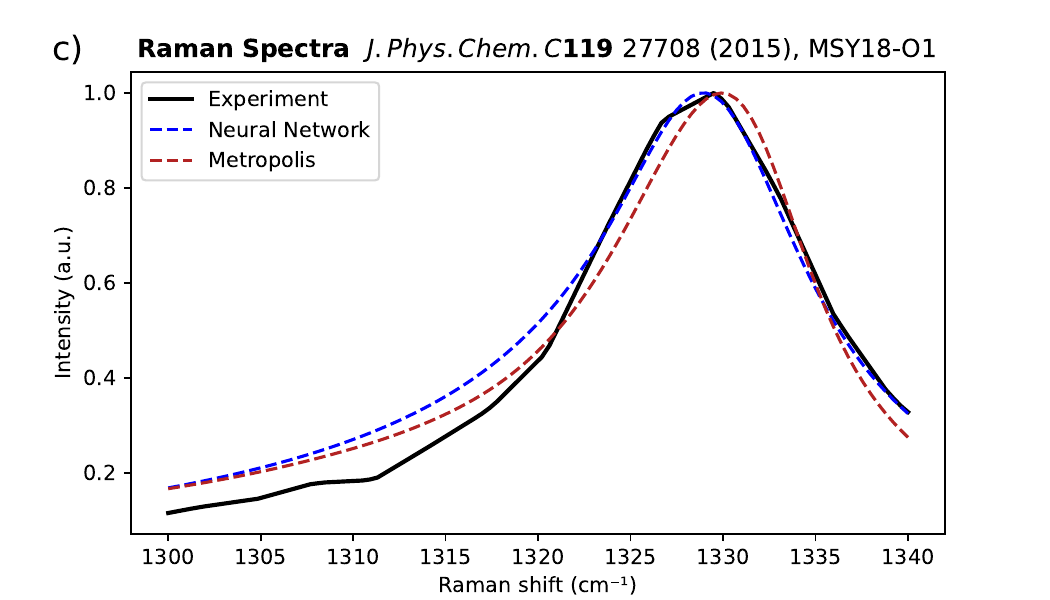}\includegraphics[width=0.49\linewidth]{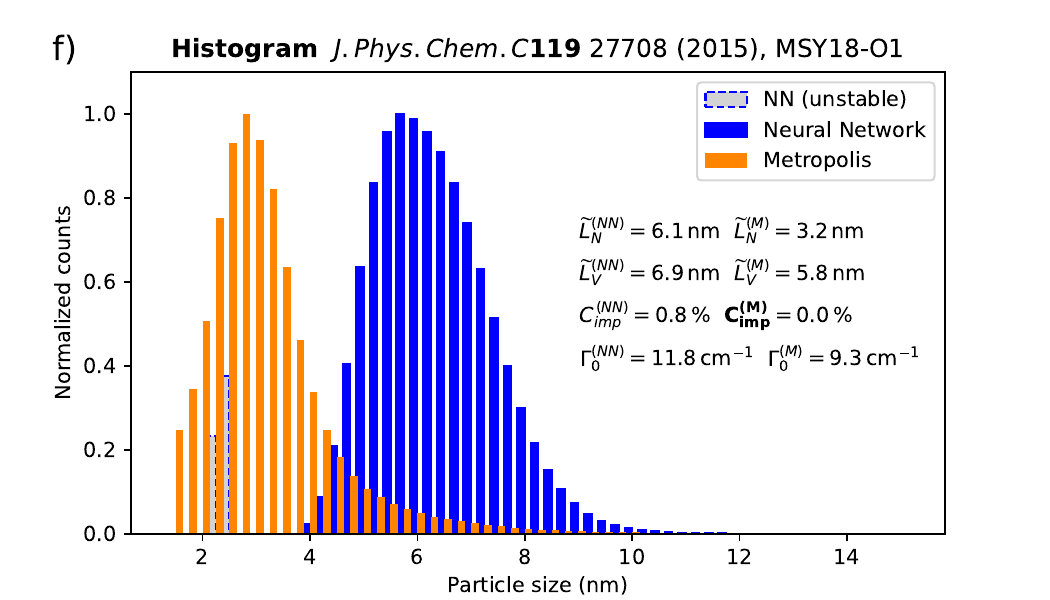} \\ 
    \caption{Here, the developed software is benchmarked on samples of various origins. Panels a) and d) are for the Raman spectra and reconstructed size distribution, respectively, for the 4.3~nm detonation nanodiamonds studied in Ref.~\cite{yoshikawa1995raman}. Panels b) and e) are for the Raman spectra and reconstructed size distribution of the adamantane HPHT sample, reported in Ref. \cite{kudryavtsev2023raman}. Finally, the Raman spectrum c) of the MSY18-O1 sample from Ref.~\cite{stehlik2015size} is analyzed with the respective size distribution shown in f). For the Metropolis fit, the defect concentration was manually set to zero, which resulted in the shift of the respective size histogram (orange bars in the histogram) to a smaller size, better coinciding with the HRTEM data. The rest of the notation follows that used in Fig.~\ref{fig_exp_shend}.}
		\label{fig_exp_misc}
\end{figure*}

Next, we studied the samples reported in various papers and obtained by different methods. Our results are summarized in Fig.~\ref{fig_exp_misc}. In the paper by Yoshikawa and coauthors dated 1995~\cite{yoshikawa1995raman}, the spectrum of nanodiamond powder obtained by detonation synthesis was reported. The powder XRD characterization revealed the nanoparticle size of 4.3~nm, which gives a reference for the size distribution obtained based on the Raman spectra. Next, we treated the spectrum from Ref.~\cite{kudryavtsev2023raman} reporting of the nanodiamonds produced by the High-Pressure High-Temperature (HPHT) treatment of pure adamantane (C$_{10}$H$_{16}$) at 12 GPa and ~1300$\degree$C. Finally, we consider the MSY18-O1 sample from Ref.~\cite{stehlik2015size}. It was mechanically ground from bulk HPHT synthetic diamond microcrystals (MSY18, initial median size 18 nm) with the subsequent annealing in air at $450\degree$C for 30 minutes and centrifugation to isolate the smallest fraction. 
%>>>>>>>>>>>>>>>>>>>>>>>>>>>>>>>>>>>>>>>>>>>>>>>>>>>>>>>>>>>>>>>>>>>>>>>>>>>>>>>>>>>>>>>>>>>>>>>
%>>>>>>>>>>>>>>>>>>>>>>>>>>>>>>>>>>>>>>>>>>>>>>>>>>>>>>>>>>>>>>>>>>>>>>>>>>>>>>>>>>>>>>>>>>>>>>>

As one sees in panels a) and d) from Fig.~\ref{fig_exp_misc}, both neural network and Metropolis approaches yield the size on the order of 3~nm, which is close to the manual fit attempt of this spectrum~\cite{yashenkin2021bench}. At the same time, the XRD gives an estimation of $\approx 5$~nm. To handle this picture, the much steeper dispersion with $B=139$~cm$^{-1}$ was proposed in the original paper~\cite{yoshikawa1995raman}. However, microscopic calculations based on the Keating model predict much less steep dispersion with $B<90$~cm$^{-1}$. In Ref.~\cite{kudryavtsev2023raman}, HRTEM evidences both small 2~nm diamond crystallites and those up to 7-8~nm. It is challenging to judge the whole histogram; however, the estimate of 5-7~nm based on Raman spectra fit does not contradict HRTEM data. For the sample  MSY18-O1 sample from Ref.~\cite{stehlik2015size}, size histograms recovered from Raman spectra overestimate size by approx 30~\%. However, if one fixes the impurity concentration $C_{imp} = 0.0$, the size distribution will perfectly coincide with the reported AFM. This example highlights the fundamental difficulty of Raman spectra interpretation, where various effects can cause similar values of the peak position redshift.

\begin{figure*}[htbp]
\centering

\includegraphics[width=0.49\linewidth]{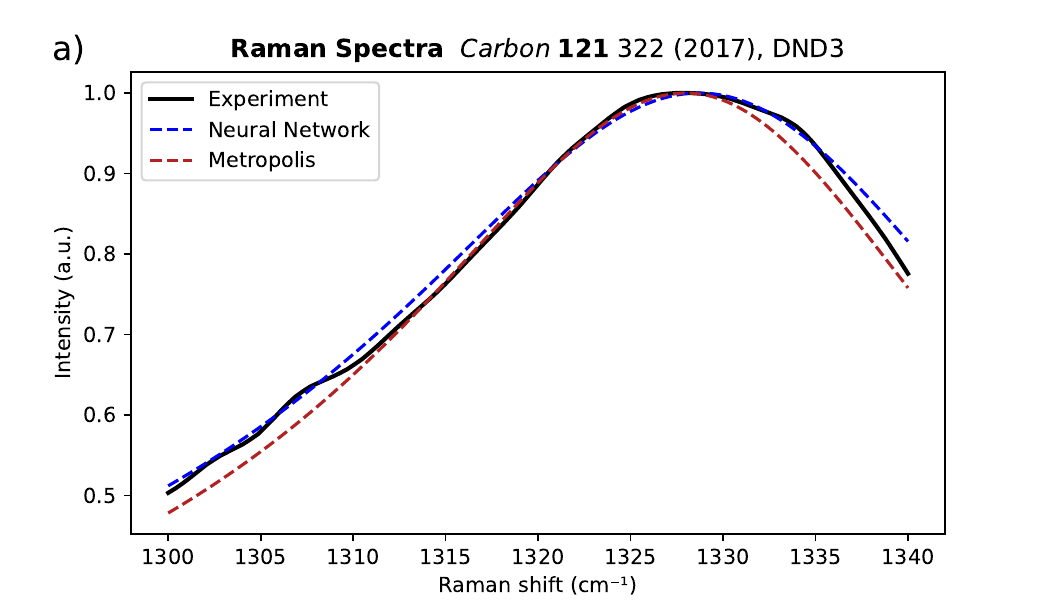}\includegraphics[width=0.49\linewidth]{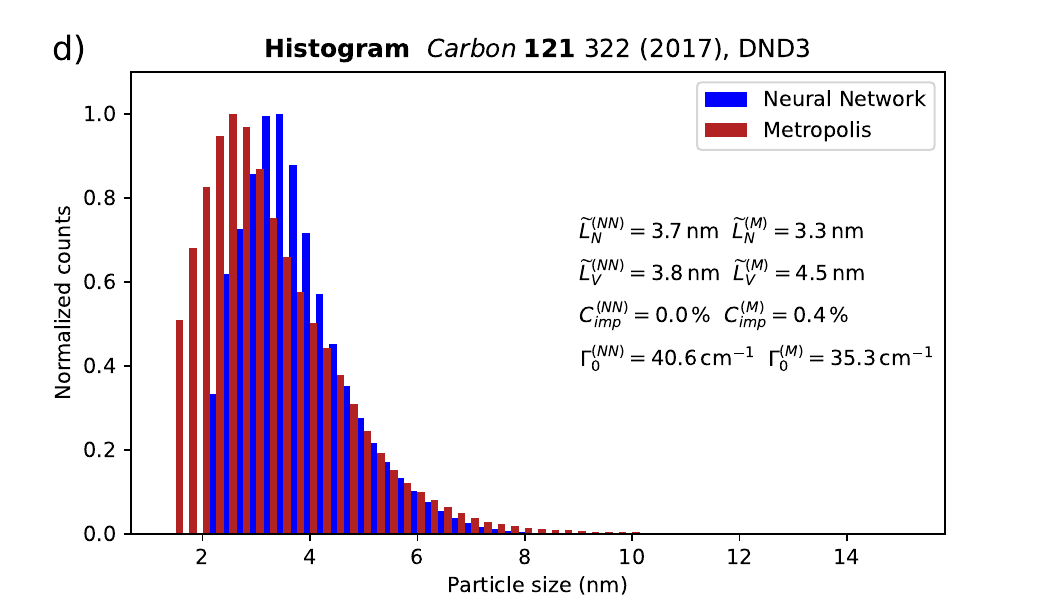} \\ \includegraphics[width=0.49\linewidth]{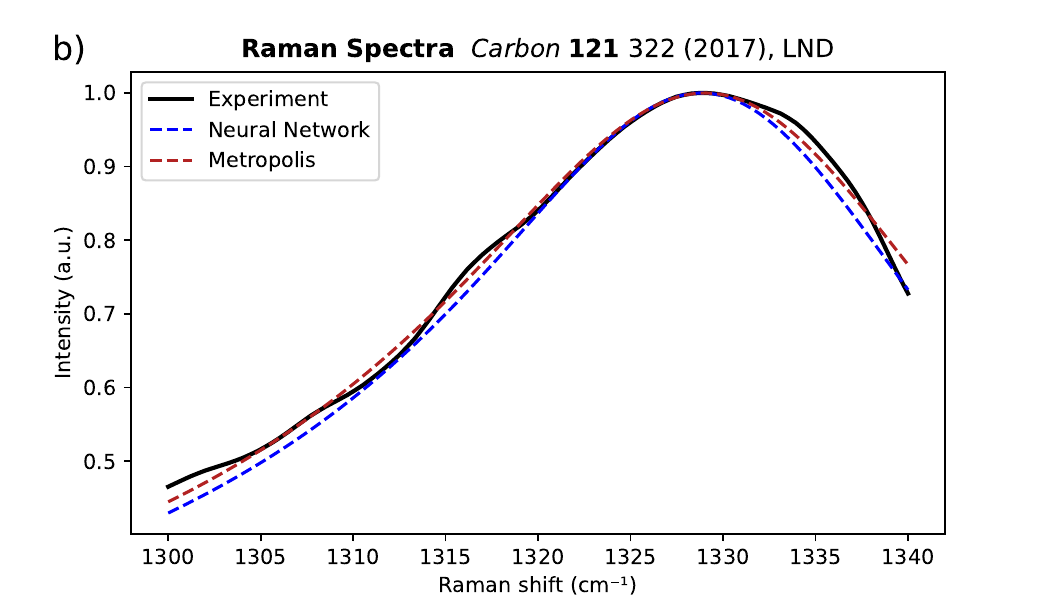}\includegraphics[width=0.49\linewidth]{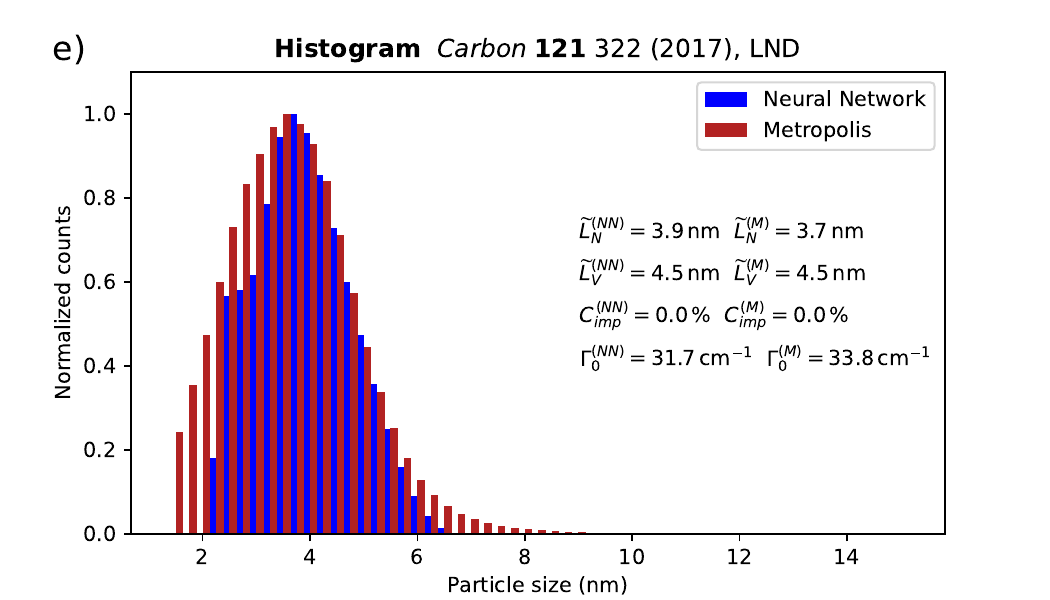} \\ \includegraphics[width=0.49\linewidth]{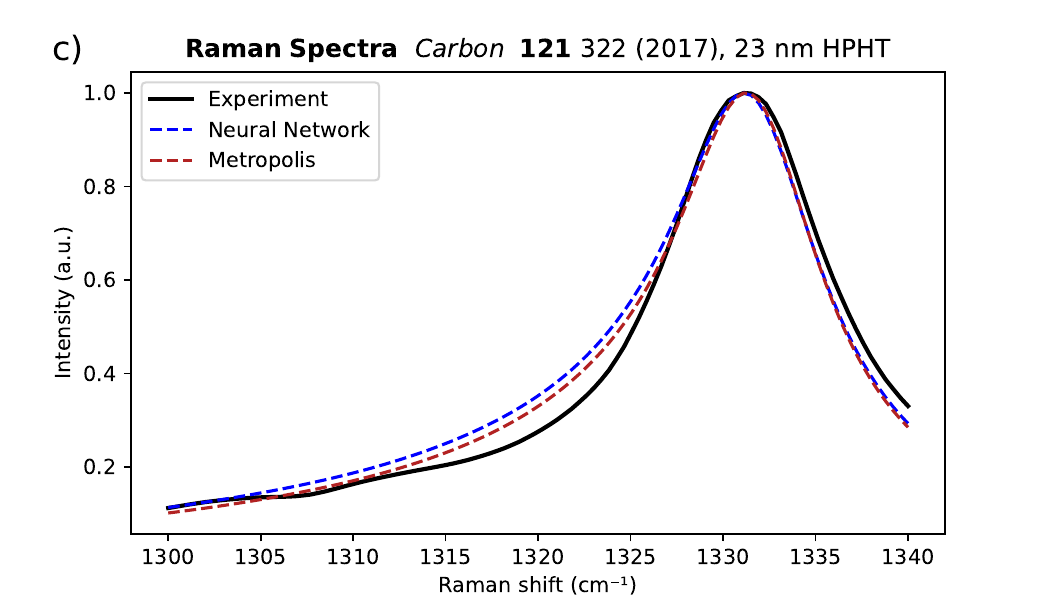}\includegraphics[width=0.49\linewidth]{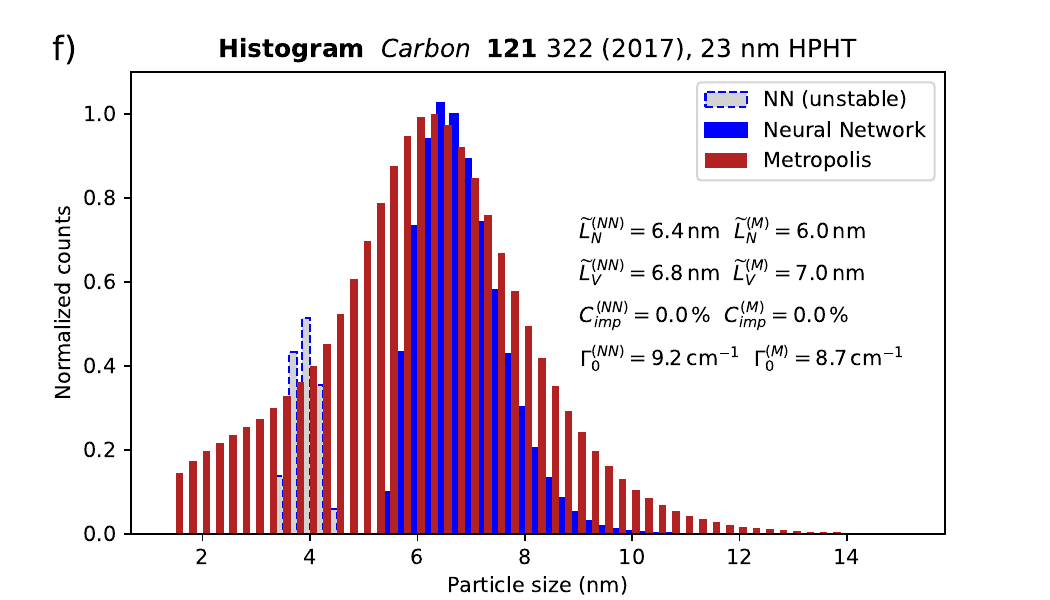} \\ 
\caption{Results of fitting of Raman spectra from Ref.~\cite{korepanov2017carbon}. Panels a), b) and c) are for the treated and reconstructed Raman spectra of the DND3, LNA, and 23~nm HPHT samples, respectively. Panels d), e), and f) are for the respective histograms. The notations are identical to those of Fig.~\ref{fig_exp_shend}.}
\label{fig_exp_carbon}
\end{figure*}

Finally, we treated Raman spectra presented in Ref.~\cite{korepanov2017carbon}, see Fig.~\ref{fig_exp_carbon}. The DND3 sample is the de-agglutinated detonation nanodiamond from NanoCarbon Research Institute Ltd., Japan, in the colloid form, with a typical size of 3.3 nm evidenced by DLS. The laser-synthesis nanodiamond (LNA)~\cite{zousman2014pure} has a 4~nm size based on XRD. Finally, the Raman spectra of the 23 nm HPHT from Microdiamant AG, acid-washed and air-oxidized, with a mean size of 23 nm according to DLS, were treated. One sees that the provided fit robustly distinguishes small (3-4~nm nanodiamonds) and those of the order of 10~nm and larger. However, for the Raman-based measurements, all sizes higher than 8~nm become practically indistinguishable.

%>>>>>>>>>>>>>>>>>>>>>>>>>>>>>>>>>>>>>>>>>>>>>>>>>>>>>>>>>>>>>>>>>>>>>>>>>>>>>>>>>>>>>>>>>>>>>>>
%>>>>>>>>>>>>>>>>>>>>>>>>>>>>>>>>>>>>>>>>>>>>>>>>>>>>>>>>>>>>>>>>>>>>>>>>>>>>>>>>>>>>>>>>>>>>>>>
	
To conclude this section, we note that the difference in characteristic size determination by both neural network-based and Metropolis methods is typically lower than 10\%. None of the method demonstrate bias in under- and overestimation of size with respect to the counterpart. Moreover, the estimations of the size-independent broadening $\Gamma_0$ by both methods are nearly similar for all treated Raman spectra of the samples of drastically differing fabrication methods. The same picture, perfect coincidence, is actual for the size-dependent broadening explained as the presence of lattice defects $C_{\textrm{imp}}$. Finally, it is pertinent to note that using the present approach, it is challenging to characterize the samples with the typical particle size larger than 8~nm and robustly determine their size distribution. It is caused by essentially low magnitude of the size quantization redshift becoming comparable with measurement accuracy for such samples, i.e., of the order of 1~cm$^{-1}$. It is clearly seen for the 23~nm sample in Fig.~\ref{fig_exp_carbon}.

\section{Discussion}

We proposed the theory of Raman spectra derivation for ensembles of diamond nanoparticles and created a ready-to-use numerical tool for its utilization. First, the theory and the numerical package allow calculating Raman spectra based on the size distribution of nanodiamonds, size-independent broadening, and knowledge of the effective lattice defects concentration contribution to the broadening and peak position redshift. Basically, the theory uses the scalar elasticity-like approximation for optical phonons that yields envelopes for atomic displacements, allowing calculation of the Raman scattering cross sections and respective phonon mode frequencies. The scattering intensity is proportional to the nanoparticle volume $\propto L^3$ and decays with principal mode quantum number as $n^{2}$ for the model case of spherical nanoparticles. The mode energy is directly the energy of a phonon experiencing the size quantization effect with the wave vector $q_n = 2\pi n / L$ corresponding to the standing wave within the nanoparticle. The good correspondence between the continuous and atomistic approaches is well validated even for acoustic phonons~\cite{combe2009acoustic}, being more sensitive to the boundary conditions. This picture is actual for optical phonons because of the vanishing amplitude at the boundaries. Thus, the effect of the nanoparticle shape is also small with respect to the redshift effect caused by general size quantization for nanopowders with significant scatter in size~\cite{utesov2018raman}. However, in the case of single-particle Raman spectra measurement or preparing the size distribution with a root mean square size deviation of the order of 10\% for 4~nm nanoparticles, the shape determination problem can be considered.

In the phonon confinement model (PCM), the continuum of phonon states from the whole Brillouin zone contributes to the Raman spectrum with the broad exponential weighting function. This broad band of phonons contributing to the Raman spectrum is, in fact, implicitly accounting for various sources of disorder: dispersion in size, lattice impurities, etc. The original formulation of PCM did not use that interpretation. Our theory considers all these effects explicitly. Constructing the spectrum within our theory is more straightforward and computationally simpler than in PCM.

Along with the spectra calculation (direct problem), the theory is useful for fitting Raman spectra with further extraction of useful information (inverse problem). First, the theory allows the creation of an extensive synthetic training dataset for the neural network. Having a Raman spectrum as an input, the trained neural network yields the vector encoding size distribution and broadening parameters. The Metropolis algorithm-inspired approach updates the system parameters step-by-step to obtain a minimal mean square deviation for the target (experimental) Raman spectrum and the theoretical one. The neural network approach allows for suppressing the human factor when fitting the spectrum. In contrast, the Metropolis-inspired approach, by manually setting the initial conditions and adding constraints on fitting parameters, is more flexible in utilizing \textit{a priori} knowledge of the nanodiamond ensemble properties. It is pertinent to note that both methods are available exclusively because of the elasticity theory-like model for optical phonon modes in nanodiamonds~\cite{utesov2018raman}.

The fitting tool was tested on 10 different nanodiamond Raman spectra, accompanied by the size measurements using complementary methods: XRD, HRTEM, or DLS. Both the neural network and the Metropolis approaches were shown to yield similar size distribution histograms, effective concentrations of defects, and size-independent broadening. In most cases, the typical nanoparticle sizes obtained from the Raman spectra fitting coincide with the data obtained by alternative methods. The practical applicability of the proposed ready-to-use software spans from the size of $\sim 2$~nm to $\sim 8$~nm. Below 2~nm, the samples have stronger surface contributions and tend to be more amorphous. Moreover, annealing is often used to reduce the diamond core and partly graphitize nanodiamonds with the formation of sp$^2$ shell~\cite{aleksenskii1997diamond,chen1999graphitization,cebik2013raman}. All that leads to sp$^3$ phase Raman signal shading by that of amorphous and sp$^2$ phase~\cite{ferrari2004raman,korepanov2017carbon,volodin2019nanocrystalline,schupfer2021monitoring}. Above 8~nm, the phonon size-quantization redshift becomes on the order of the Raman spectra measurement precision~\cite{koniakhin2018raman}, thus being an unreliable size marker. Importantly, the effect of disorder also contributes to the Raman peak position redshift. So, it acts in the same way as the decrease in the typical size of nanoparticles. Sometimes these contributions are hard to distinguish, which is illustrated by one of the treated Raman spectra. However, based on the \textit{a priori} knowledge of the defectiveness of the sample, one can fix $C_\text{imp}$ to perform the Metropolis method. If information on the defectiveness from Raman measurements is necessary, the usage of the neural network approach is preferable. The obtained values of effective concentration of lattice impurities are comparable with the non-carbon atoms concentration~\cite{shenderova2011nitrogen, osipov2019nitrogen, stehlik2021size_and_nitrogen} and far higher than that of NV-centers~\cite{chang2016counting,osipov2019nitrogen}.

\section{Conclusion}

In summary, we present a simple, ready-to-use set of instruments to work with the Raman spectra of nanodiamonds. It allows solving both the forward problem of spectrum calculation and the inverse problem of spectrum fit and extraction of nanodiamond parameters (size and defectiveness). For the latter, we provide two tools, based on a machine learning paradigm: a neural network approach and the Metropolis algorithm. Their identical capability of reconstructing nanodiamond size distribution functions is shown using various experimental Raman spectra reported in the literature.

\section*{CRediT authorship contribution statement}

Sergei V. Koniakhin:  Conceptualization, Theory, Data curation, Investigation, Formal analysis, Software backend, Software frontend,  Writing original draft. Oleg I. Utesov: Theory, Formal analysis, Software test, Writing and editing. Vitaly I. Korepanov: Software test, Writing and editing. Andrey G. Yashenkin: Theory.

\section*{Data availability statement}

The data supporting the conclusions of this study are available from the corresponding author upon reasonable request.

\section*{Declaration of competing interest}

The authors declare that they have no known competing financial interests or personal relationships that could have appeared to influence the work reported in this paper.

\section*{Acknowledgements}

S.V.K. and O.I.U. acknowledge financial support from the Institute for Basic Science (IBS) in the Republic of Korea through YSF Project No. IBS-R024-Y3 O.I.U. acknowledges Project No. IBS-R024-D1. V.K. was supported by the State Task \textnumero075-00295-25-00.
    
	\bibliographystyle{unsrt}
	\bibliography{ref.bib}
	
\end{document}